\documentclass{article}
\usepackage{graphicx}

\begin{document}

\leftline{\it in
  A.W. Alsabti, P. Murdin (eds.), Handbook of Supernovae}
\leftline{\it Springer International Publishing AG 2016}
\leftline{DOI 10.1007/978-3-319-20794-0-16-1} 

\vskip1cm

\begin{center}
{\Large \bf Gould's Belt: Local Large Scale Structure in the Milky Way} 
\vskip0.5cm
{\large Jan Palou\v{s} \& So\v{n}a Ehlerov\'{a}}
\vskip0.3cm
{\it Astronomical Institute, Academy of Sciences of the Czech Republic, Bo\v{c}n\'{i} II 1401/1, Prague, Czech Republic}
\end{center}
\vskip1cm

\begin{abstract}
Gould's Belt is a flat local system composed of young OB stars, molecular clouds and neutral hydrogen within 500 pc from the Sun. It is inclined about 20 degrees  to the galactic plane and its velocity field significantly deviates from  rotation around the distant centre of the Milky Way. We discuss possible models of its origin:  free expansion from a point or from a ring, expansion of a shell, or a collision of a high velocity cloud with the plane of the Milky Way.  Currently, no convincing model exists. Similar structures are identified in HI and CO distribution in our and other nearby galaxies. 
\end{abstract}
  
\section{Introduction}
Young OB stars, interstellar dark clouds, molecular clouds, dust  and neutral HI in the local volume within 500 pc of the Sun form a flat system tilted relative to the Galactic plane by about 20 degrees, it is called the Gould's Belt.  Velocity field of young stars in the Gould's Belt distinctly deviate from the rotation around the distant centre of the Galaxy: they show systematic expansion and additional rotation in the same sense as the galactic  rotation, but, around a centre that is much closer to the Sun than is the centre of the Milky Way. Origin and evolution  of the Gould's Belt is explored, however, currently convincing model does not exist.  HI supershells discovered in observations show that there may be many similar structures in the Milky Way and other nearby galaxies.   Reviews on the Gould's Belt has been published by P\"{o}ppel (1997) and Palou\v{s} and Ehlerov\'{a} (2015).

\section{Observations}
\subsection{Space  Distribution}

The space distribution of young OB stars has been analyzed by Stothers and Frogel (1974), Palou\v{s} (1983, 1985),  Westin (1985), Comeron et al. (1992, 1994) and others:  young stars are concentrated in two distinct planes, galactic and Gould's Belt. Young stars in Scorpius-Centaurus, Orion Perseus, Vela, Scutum and Vulpecula form a flat system tilted relative to the Galactic plane by about 20 degrees with the ascending node in the direction l = 295 degrees which reaches about 200 pc above the galactic plane in the direction to the galactic centre and 500 pc below the galactic plane away the galactic centre.  OB associations (Blaauw 1985, 1991)  show a local hole of diameter of about 200 pc. The dark clouds (Sandqvist, 1977),  molecular clouds (Blitz et al. 1984) and local HI (Lindblad, 1967 -- feature A) follow a belt-like space distribution surrounding the local hole in the interstellar medium.  The schematic picture of the distribution of young stars and interstellar matter  was sketched by Elmegreen (1992), see Fig 1.

\begin{figure*}
\includegraphics[width=0.9\textwidth]{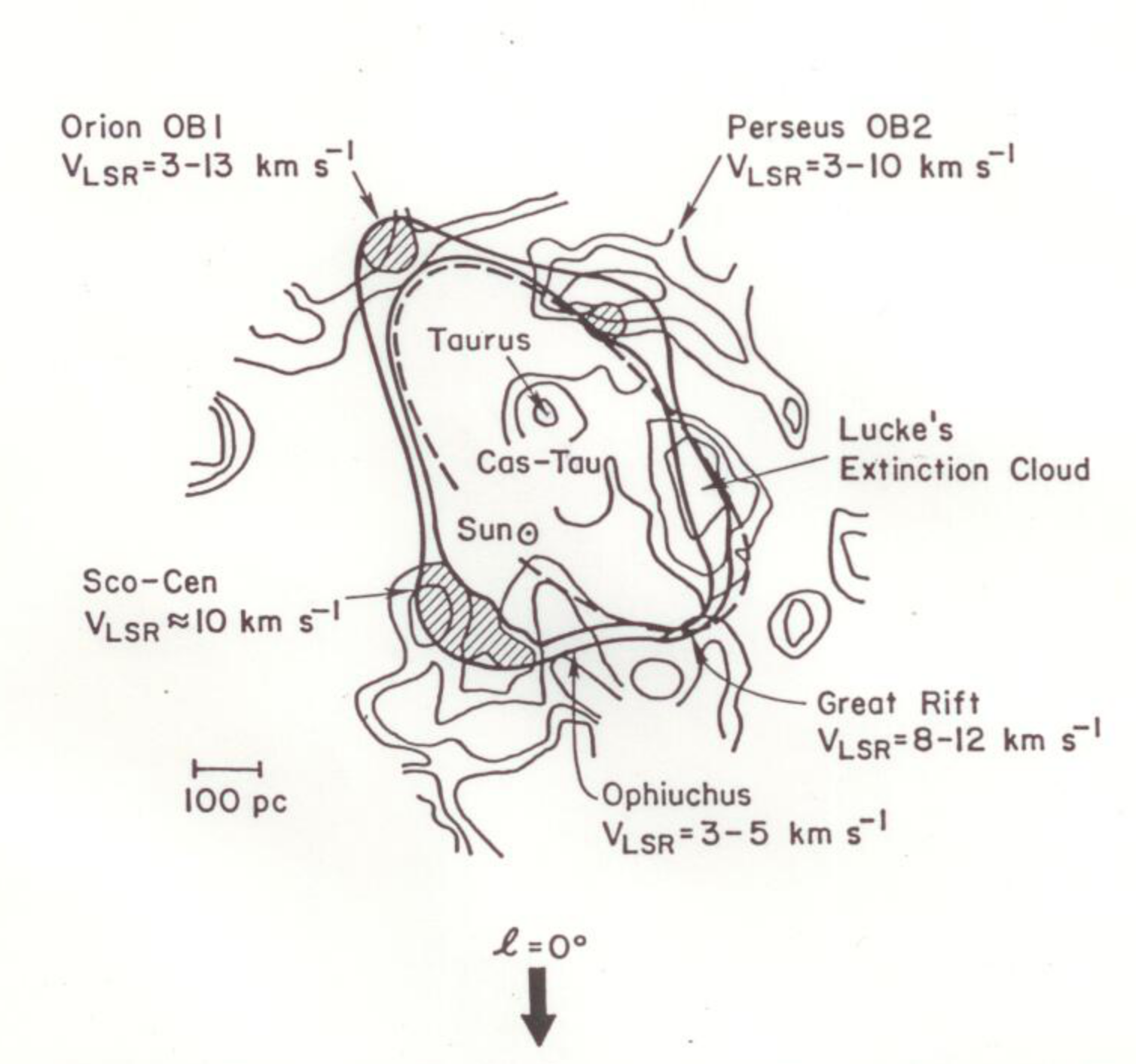}
\caption{The picture by Elmegreen (1992) of local clouds and OB associations forming the Gould's Belt. The dashed line shows the Lindblad et al.  (1973)  solution of freely expanding ring.}
\label{fig1}
\end{figure*}

\subsection{Kinematics}

The stellar velocity field in the Solar vicinity is interpreted as due to rotation around the distant galactic centre in the direction of the constellation Sagittarius.  Radial velocities and proper motions may be described by two  Oort constants A and B, that are the two terms of nine in the linear approximation of the flow near the Sun. It assumes a smooth velocity field in the solar vicinity without any abrupt ripples and jumps with radial and tangential velocities corresponding to the galactic differential rotation, i.e. motions perpendicular to the line Sun – centre of the Milky Way. The values A = 13 km/s/kpc and B = -13 km/s/kpc show that the rotation curve of the Milky Way  in the solar vicinity is flat. The absence of all other linear terms that proves that velocities reflect the rotation around the centre of the Milky Way parallel to the galactic plane.
However, the local young objects show more complex velocity pattern.  In the region of the Gould's Belt the velocity field differs (Lindblad et al., 1997): A is positive, and B is deeply negative, and also other constants describing the velocity field are non-zero. The Gould's Belt shows radial expansion from a local centre in the direction of Cas-Tau association.  It also shows the rotation around the local centre in the same sense as the rotation around the distant centre in Sagittarius (Fig. 2).  
Using positions and space velocities of young stars of the Gould's Belt, we trace their orbits in the Milky Way. The smallest volume is occupied 10 – 12 Myr ago when  the Gould' Belt progenitors reside in a sheet-like region about 500 pc long and 100 pc broad with the main axis pointing in the direction l = 20 -- 200 degrees (Palou\v{s}, 1998a, b). Such sheet may be formed in collision between two supershells. This model needs to be explored in the future.
In the direction perpendicular to the galactic plane, the average motion is proportional to the distance from the rotational axis pointing to the direction  l = 160 -- 340 degrees (Comeron, 1999). Gould' Belt shows the coherent motion in the direction perpendicular to the Galaxy plane, however, its kinematical axis is displaced from the line of nodes of the Gould' Belt.    

\section{Models}

Models of free expansion from a small region, or expansion of HI supershell, its fragmentation and star formation in homogeneous or porous medium distributed in the galactic plane or in the inclined gas layer show a velocity field that does not resemble observations. Alternative models of a collision of a high velocity cloud with the galactic HI plane that may fuel the Gould's Belt expansion  do not offer an explanation of the inclination  of its plane relative to the galactic plane. New models based on a collision between two HI supershells need to be explored.

\begin{figure*}
\includegraphics[width=0.9\textwidth]{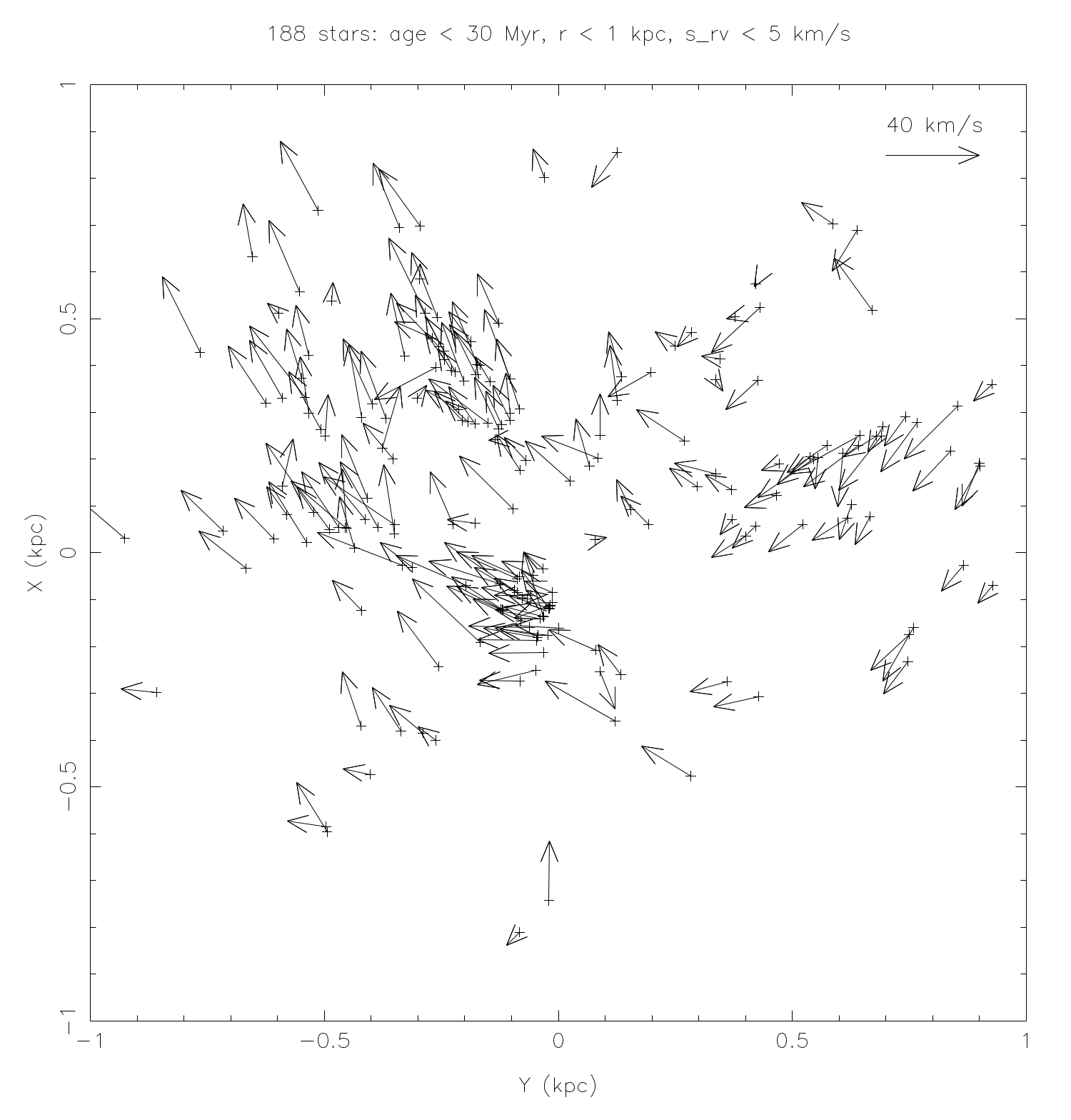}
\caption{The components of the space velocity vectors parallel to the galactic plane of the young OB stars within 1 kpc from the Sun.}
\label{fig2}
\end{figure*}

\subsection{Free Expansion}

Blaauw (1952) introduced a model of an expanding stellar association. Later, this free expansion model was used  by Lindblad et al. (1973) to interpret the kinematics of the HI ring around the Sun. To fit the size and radial velocity distribution, the initial expansion velocity needs to be a few km/s only and the expansion time  close to 60 Myr. The expansion shown by young OB associations and HI may be reproduced, however, the large and negative B known from the analysis of tangential velocities  of Gould' Belt young stars is not: all expanding models show B close to zero, very different than the value derived from observations.

\subsection{Expanding shells}

Olano (1982) proposes an expanding supershell that acquires its energy from stellar winds and supernovae of an OB association. A similar 3-dimensional model of a shell expanding as a snow-plough in the ISM of the Galaxy was examined by Ehlerov\'{a} et al. (1997). Different variants of the 3D expanding supershell were analysed  by Perrot and Grenier (2003), who included abrupt or gradual energy injection, accretion of the interstellar medium, fragmentation of the shell and porosity of the ambient medium, or star formation in pre-existing clouds. None of these models is able to explain low and negative value B as derived from motions of young stars, also the inclination of the Gould's Belt relative to the galaxy plane remains unexplained.

\subsection{Impact of a high velocity cloud}

Tenorio-Tagle (1980) proposes an impacts of a high velocity clouds into the HI galactic disk as a possible explanation of the observed HI supershells of sizes hundred of pc or more. Kinetic energy involved in formation of an expanding supershell is similar to the kinetic energy of a high velocity cloud and also size of the region  and final expanding velocity after a collision correspond to an observed supershell (Tenorio-Tagle et al., 1986). Gould's Belt may be formed due to fragmentation of an expanding HI supershell formed after a high velocity cloud collided with the galaxy HI plane. However, also in this model, the inclination of the Gould's Belt relative to the galaxy plane remains unexplained.

\subsection{Other Gould's Belts}

HI distribution in the Milky Way (Ehlerov\'{a} and Palou\v{s}, 2013) and in other galaxies (Bagetakos, 2011) shows large scale structures, supershells. They are correlated with molecular clouds , which frequently cluster along their rims (see also Ehlerov\'{a} and Palou\v{s}, 2016). An example of the large HI structure GS24-03+37 is shown in Fig. 3. Its size (nearly 1 kpc) is similar to the size of the Gould's Belt. We conclude that there is many HI -- CO -- young stars structures in galaxies. Gould's Belt is just the local example of the ISM -- SF connection.

\subsection{Gas flows and future models}

Models of the gas flows and molecular cloud structure formation in the galactic potential with the spiral arms (Dobbs, 2015) demonstrate that there may be many structures similar to the Gould's Belt caused by the curling flows after the passage of the galaxy spiral arm. Their shapes only partially depend on stellar feedback, the shapes follow gas flows in the large scale galactic potential. Some of these structures display voids and shapes similar to the Gould's Belt. Future models should include galactic large-scale atomic and molecular gas flows in a model of the galaxy with energy and momentum feedback from young OB associations that form in places of molecular cloud formation. Eventual interaction of expanding supershells should be explored in order to see if the fragmentation and synchronized star formation can be achieved.

\section{Conclusions}

The Gould's Belt serves as an example of the local star formation correlated at the kpc scale. This large-scale  star forming region is identified in spatial distribution of young stars and ISM, and by their velocity distinct velocity field that deviates from circular rotation around the galaxy centre. It seems that there are many structures similar to the Gould's Belt in HI and CO in the Milky Way and other nearby galaxies. Such structures may result from a combination of the galactic gas flows and stellar feedback. More models on neutral gas flows, molecular cloud formation, star formation and energy and momentum feedback from young stars should be explored.

\begin{figure*}
\includegraphics[width=0.9\textwidth]{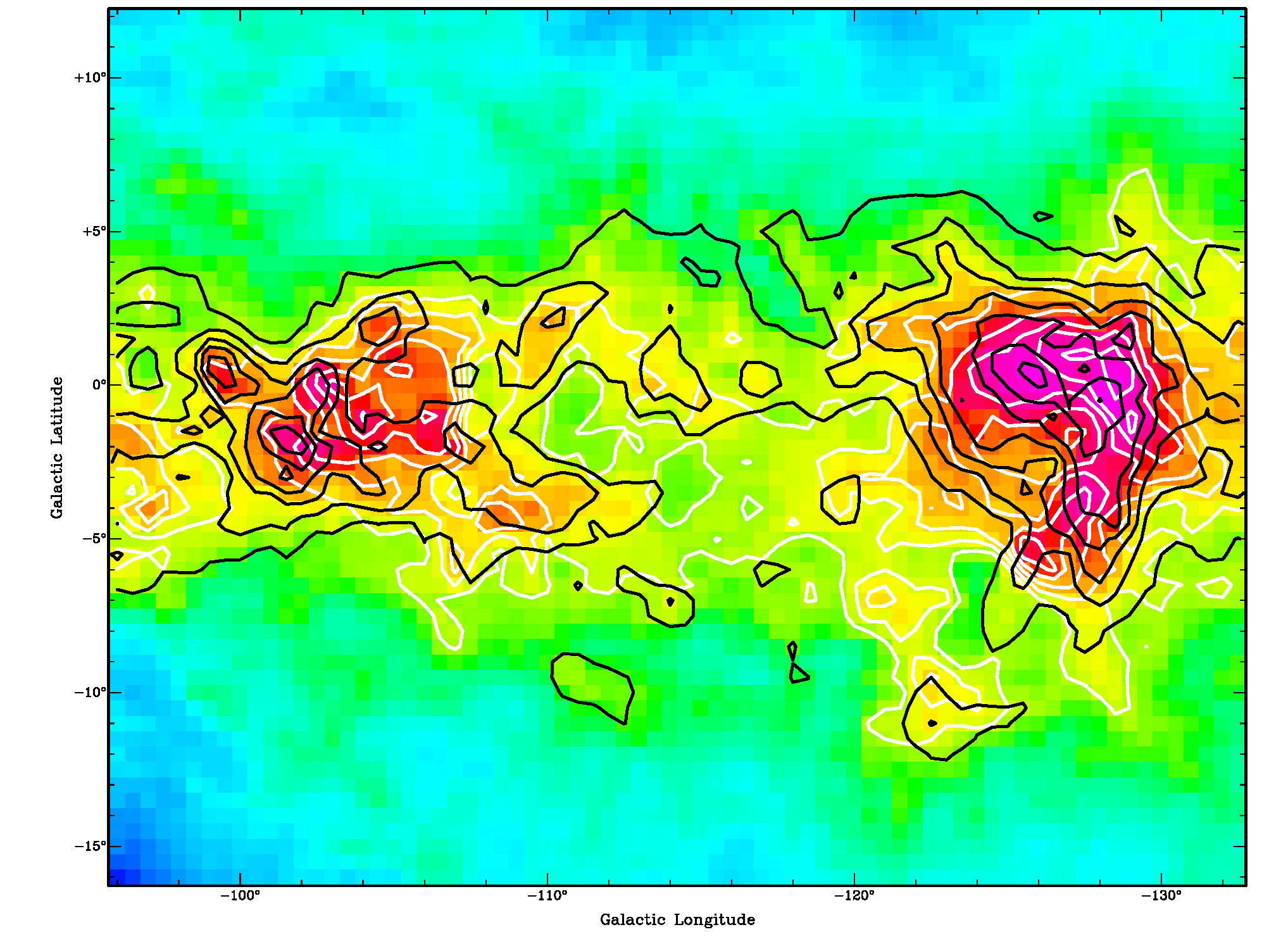}
\caption{GS24-03+37: the HI column density (map) with contours of HI integrated over the the closer half of the structer (white contours) and  over the farther half (black contours). The shift between white and black contours indicate a ring of a denser gas tilted towards the Galactic plane. 10 degrees correspond to about 700 pc.}
\label{fig3}
\end{figure*}

\subsubsection*{Acknowledgements}

This study has been supported by Czech Science Foundation grant 209/12/1795 and by the project RVO: 67985815.

\subsection*{References}

Bagetakos, I., Brinks, E., Walter, F. et al., 2011, AJ, 141, 23 \\
Blaauw, A., 1952, BAN, 11, 414 \\
Blaauw, A., 1985, IAU Symp. 106, 336 \\
Blitz, L.,  Magnani, A. and Mundi, L., 1984, ApJ, 282, L9   \\
Blaauw, A., 1991, in “Physics of Star Formation and Early Stellar Evolution”, eds N., Kylafis and Ch., Lada, Kluwer Acad. Pub., p. 125  \\
Comeron, F., 1999, A\&A, 351,506 \\
Comeron, F., Torra, J. and Gómez, A., 1992, Ap \& SS, 187, 187 \\
Comeron, F., Torra, J. and Gómez, A., 1994, A\&A, 286, 789 \\
Ehlerov\'{a}, S. and Palou\v{s}, J., 2013, A\&A, 550, 23 \\
Ehlerov\'{a}, S. and Palou\v{s}, J., 2016, A\&A, 585, 5 \\
Ehlerov\'{a}, S., Palou\v{s}, J., Theis, Ch. and Hensler, G., 1997, A\&A, 328, 121 \\
Elmegreen, B. G., 1992, in ”Evolution of Interstellar Matter and Dynamics of Galaxies”, eds. J., Palou\v{s}, W. B. Burton and P. O. Lindbad, Cambridge Univ. Press, p. 178  \\
Gould, B. A., 1874, Proc. Amer. Assoc. for Adv. Science p. 115 \\
Gould, B. A., 1879, Uranometria Argentina, Result. Obs. Nac. Argentino, I. p. 354  \\
Herschel,  J.F.W., 1847, Results of Astron. Observations made during the years 1834 – 1838 at the Cape Good Hope, London, p. 385 \\
Klepesta, J. and Ruckl, A., 1960, Colour Map of the Southern Sky 1950.0, Central Office of Geodesy and Cartography, Prague \\
Lindblad, P. O., 1967, Bull. Astron. Inst. Netherlands, 19, 34  \\
Lindblad, P. O.,  Grape, K., Sandqvist, Aa. and Schober, J., 1973, A\&A, 24, 309  \\
Lindblad, P.O., Palou\v{s}, J., Lodén, K. and Lindegren, L., 1997, Hipparcos Venice' 97, ESA,  p. 507 \\ 
Olano, C. A., 1982, A\&A, 112, 195 \\
Palou\v{s}, J., 1983, Bull. Astron. Inst. Czechoslov., 34, 286 \\
Palou\v{s}, J., 1985, Bull. Astron. Inst. Czechoslov., 36, 261 \\
Palou\v{s}, J., 1998a, in N. Capitaine and J. Vondrák (eds), Journeé systemes de reference spatiotemporel 1997, p. 157 \\
Palou\v{s}, J., 1998b, in W. J. Duschl and Ch. Einsel (eds), Dynamics of Galaxies and Galactic Nuclei, ITA Proceeding Series, Vol. 2, p. 157 \\
Palou\v{s}, J., Ehlerov\'{a}, S., 2015, in Lessons from the Local Group, K. Freeman et al. (eds.), Springer, p. 63 \\
P\"{o}ppel, W., 1997, Fundamentals of Cosmic Physics, 18, 1 \\
Sandquist, Aa., 1977, A\&A, 57, 467 \\
Stothers, R. and Frogel, J. A., 1974, Astron. J. 79, 456 \\
Tenorio-Tagle, G., 1980, A\&A, 88, 61 \\
Tenorio-Tagle, G., Bodenheimer, P., Rozyczka, M. and Franco, J., 1986,  A\&A, 170, 107 \\
Westin, T. N. G., 1985, A\&AS, 60, 99 \\

\end{document}